\documentclass[11pt,titlepage,final]{article}

\usepackage{graphicx}
\usepackage[authoryear]{natbib}
\newcommand{\etal}{{\em et al.}}

\newcommand{\kms}{\rm{km}\,{\rm{s}}^{-1}}
\newcommand{\kpc}{\rm{kpc}}
\newcommand{\kmskpc}{\rm{km}\,{\rm{s}}^{-1}\,{\rm{kpc}}^{-1}}
\begin{document}

\title{Milky Way Gas Dynamics\thanks{Invited talk at the AAS Division on
    Dynamical Astronomy meeting, Santa Barbara, April 2005}}
\author{Peter Englmaier \\
Peter.Englmaier@unibas.ch \\
+41-61-2055-434\\
\\
Departement f\"ur Physik und Astronomie,\\
Astronomisches Institut, Universit\"at Basel,\\
Venusstr. 7, 4102 Binningen, Switzerland\\
\and Ortwin Gerhard \\
gerhard@mpe.mpg.de\\
+41-61-30000-3539\\
\\
Max-Planck-Institut f\"ur Extraterrestrische Physik,\\
Giessenbachstrasse, 85748 Garching bei M\"unchen, Germany\\}
\maketitle
\begin{abstract}
The Milky Way is made up of a central bar, a disk with embedded spiral
arms, and a dark matter halo. Observational and theoretical
constraints for the characteristic parameters of these components will
be presented, with emphasis on the constraints from the
dynamics of the Milky Way gas. In particular, the fraction of dark
matter inside the solar radius, the location of the main resonances,
and the evidence for multiple pattern speeds will be discussed.
\end{abstract}

\section{Introduction}

In our current understanding of the universe galaxies evolve within
dark matter halos during the expansion of the universe.  Very
successfully, the theory of hierarchical clustering and merging of
dark matter halos explains the structure on large scales and the
variety of galaxy masses and types observed. Yet, on smaller scales
the details of dissipation, star formation, and secular evolution,
which also depend on non-gravitational forces, become more important
and need to be addressed.  Observations and detailed modeling is
required to help understand the processes of galactic evolution.

Our own galaxy is particularly important in this respect, because it
allows much more detailed observations, down to stellar proper motions
around the central black hole and planet formation around young
stars. Many new large surveys of the visible matter inventory of the
Milky Way are under way: the survey RAVE hopes to measure radial
velocities and chemical abundances of the brightest 50 million stars,
the space interferometry mission SIM will accurately determine the
distance to the galactic center and the local standard of rest (LSR)
velocity, and GLIMPSE, a Spitzer Legacy Science Program, already built
a catalog of 47 million sources in the longitude range
$10^\circ\leq|l|\leq 64^\circ$ within the galactic plane $|b|\leq
1^\circ$, to be extended to fill the gap within $|l|\leq
10^\circ$. The goal of these surveys is to understand the Milky Way's
present state and formation history \citep{GLIMPSE,SIM,RAVE}.

Detailed modeling of the Milky Way is required to make sense of these
data sets. Because of our position within the galactic plane, we
cannot grasp the large-scale structure easily.  With certain symmetry
assumptions, however, it is possible to recover the present
3-dimensional structure. In the following, we review the
steps necessary for deriving such a model.

\section{Luminosity model}

Traditionally, mass models for the Milky Way have been constructed
from parametrized analytical descriptions of the individual
components: bulge, thin and thick disk, bar, and halo.  The most
successful axisymmetric model was published by \citet{Kent1992} who
used IRAF data to describe the stellar matter distribution. Later
\citet{Dwek++1995} used the COBE/DIRBE near-infrared maps of the inner
Galaxy to fit a tri-axial bulge and double-exponential disk to the
near-IR data. This approach was so successful --- for the first time
the bar of the Milky Way was directly seen --- that a number of more
detailed studies were triggered.  However, any division into
components is to some degree arbitrary, and parametric models have
serious limitations.

Another route was taken by \citet{Fux1997}, who
generated a large number of bar forming n-body models. By selecting
the model closest to the observed COBE/DIRBE light distribution he was
able to reproduce the observations.  The method is limited by the 
number of models one
can generate; and there is no guarantee that and how well the Milky
Way fits in the space spanned by the n-body models evolved
from the chosen initial conditions.

A more generic approach was adopted by \cite{BinneyGerhard1996} and
\cite{BissantzGerhard2002}, who developed and applied new
non-parametric methods, to fit the observed near-IR light distribution
as measured by COBE/DIRBE.  Both methods require the specification of
an initial model which is subsequently adjusted to simultaneously
minimize the deviations from the observations, smoothness and penalty
term.  With some reasonable symmetry assumptions, these methods yield
the approximate 3-dimensional light distribution of the inner Galaxy
on a model grid which can then be analyzed to extract parametric
information, such as disk scale length or bar axis ratio and radial
extent.

When adding a gaseous component, the bar in this
model can drive spiral arms similar to the observed spiral pattern.
However, a more realistic treatment of the spiral arms was introduced
by \cite{BissantzGerhard2002} who allowed for 
deviations from triaxial symmetry, and added an analytical spiral arm model
known to fit the observed spiral arm pattern rather well, and took 
it into account when 
inverting the light distribution. Interestingly, a model with
spiral arm contribution to the COBE/DIRBE light makes the deduced 
bar longer and more elongated compared to models without spiral
arms. Furthermore, only a model with spiral arm contribution can
produce the observed asymmetries in the red clump line-of-sight
distributions \citep{BissantzGerhard2002}. This result demonstrates
how important it will be to combine different data sets from different
observations.

The modeling results depend to some extent on the assumed solar
galactocentric radius $R_\odot$ and local standard of rest (LSR)
velocity $V_\odot$. \citet{Reid1993} reviewed measurements of
$R_\odot$ and found $8\pm0.5\,\kpc$, while more recently
\citet{Eisenhauer++2003} estimated the distance to the galactic center
black hole to be $7.94\pm0.42\,\kpc$, still with a relative error of
$5\%$. The LSR velocity is then inferred from Hipparcos data, coming
out to be $218\pm21\,\kms$ \citep{FeastWhitelock1997}.  Most of the
uncertainty comes from the $R_\odot$ error. A new analysis, taking
into account previously ignored systematic effects, was published by
\citet{Kalirai++2004} showing that the LSR velocity might be lower,
only $203\pm25\,\kms$. \citet{ReidBrunthaler2004} on the other hand
found a higher value from the proper motion of Sgr~A$^*$:
$V_\odot=236\pm15\,\kms$.  Here we assume a solar galactocentric radius
$R_\odot=8\,\kpc$ and a LSR velocity $V_\odot=220\,\kms$.

\section{Mass model}

Assuming a spatially constant mass-to-light ratio, the NIR luminosity
model can be converted to a mass model.  In reality, the
mass-to-light ratio of a stellar population depends on age and 
metallicity of the underlying stellar population, which would vary
within the Milky Way, but these dependencies are weaker than in the
optical. The luminous Galaxy is probably embedded in a dark matter
halo, which makes the circular rotation curve flat in the outer parts.
Combining the luminosity model and the constant $M/L$ assumption for the
stars with an analytical model for the dark matter distribution, a
mass model can be constructed with only a few free parameters: the mass-to-light
ratio for the visible component(s), the position angle of the bar, 
the pattern speeds of bar and spiral pattern, and the halo model 
parameters.

For the invisible dark matter halo, we assume a logarithmic density
profile with two parameters: scale radius $a$ and velocity $V$. We
find below that the details of the assumed profile do not matter, as long
as the rotation curve of the dark halo is proportional to radius
within the region probed by the data, i.e. inside $R_\odot$. 
Therefore, the dark matter halo model truly has only one free
parameter, the slope at small radii $V/a$, where $a$ can be anything
larger than $~10\,\kpc$.

\begin{figure}
\includegraphics[width=\textwidth]{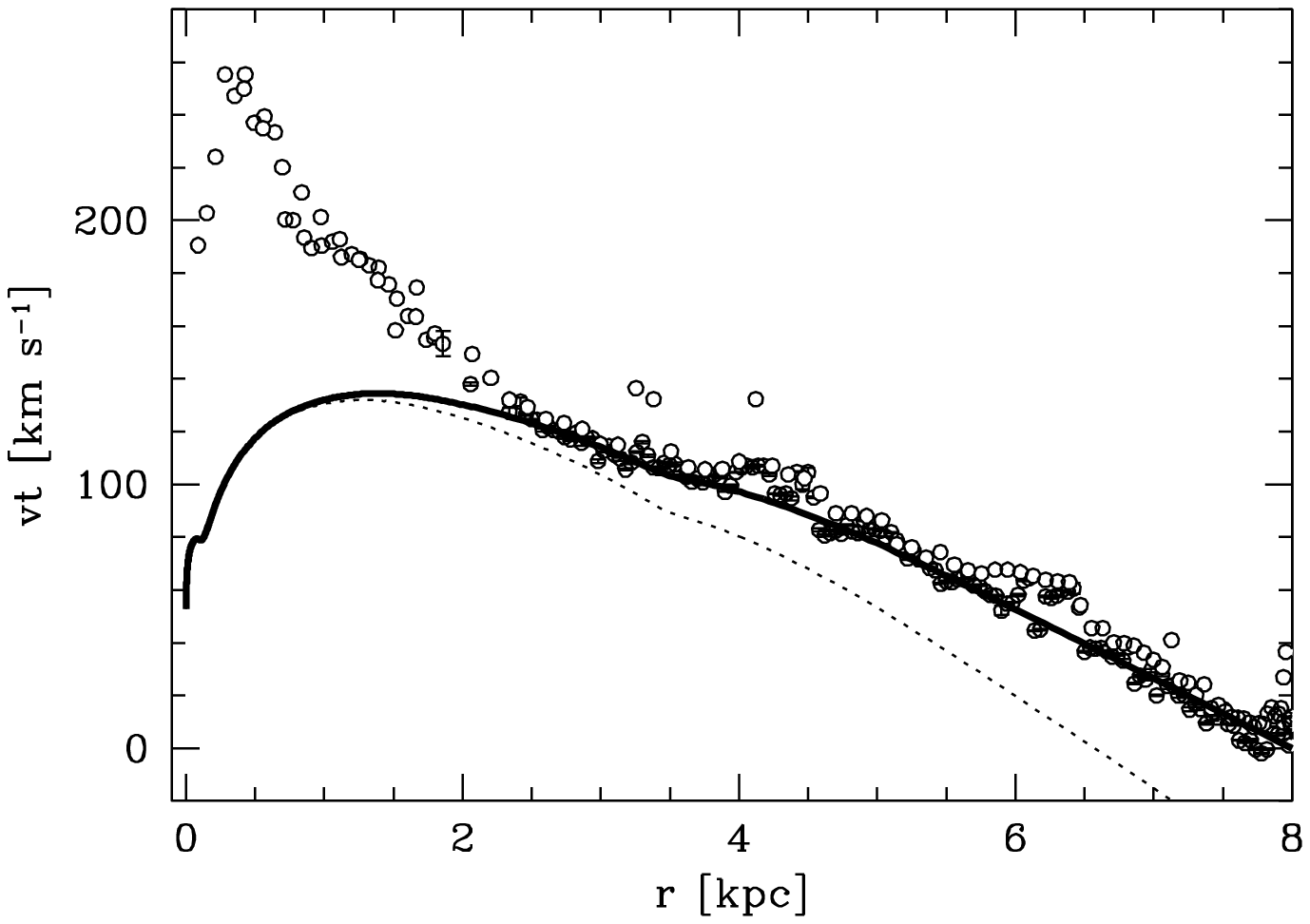}
\caption{Comparison of model (solid curve) to a
  collection of terminal velocity curve data (symbols: $^{12}$CO data
  from \citet{Clemens1985}, and HI data from \citet{BurtonLiszt1993,Fich++1989}). 
  The lower dotted curve corresponds to the visible light contribution. For this fit,
  we assume $R_\odot=8\,\kpc$ and solve for the LSR velocity. The fit
  is required to match zero terminal velocity at $R_\odot$ and the only
  free parameters varied are the halo velocity scale $V$ and the
  mass-to-light  ratio for the visible matter, which depends weekly on
  the assumed $V$. By adjusting $V$ we were able to match the observed
  terminal velocity curve; the LSR velocity for this fit is $V_\odot=221\,\kms$, in
  agreement with observations.
  \label{talk3}
}
\end{figure}

\begin{figure}
\includegraphics[width=\textwidth]{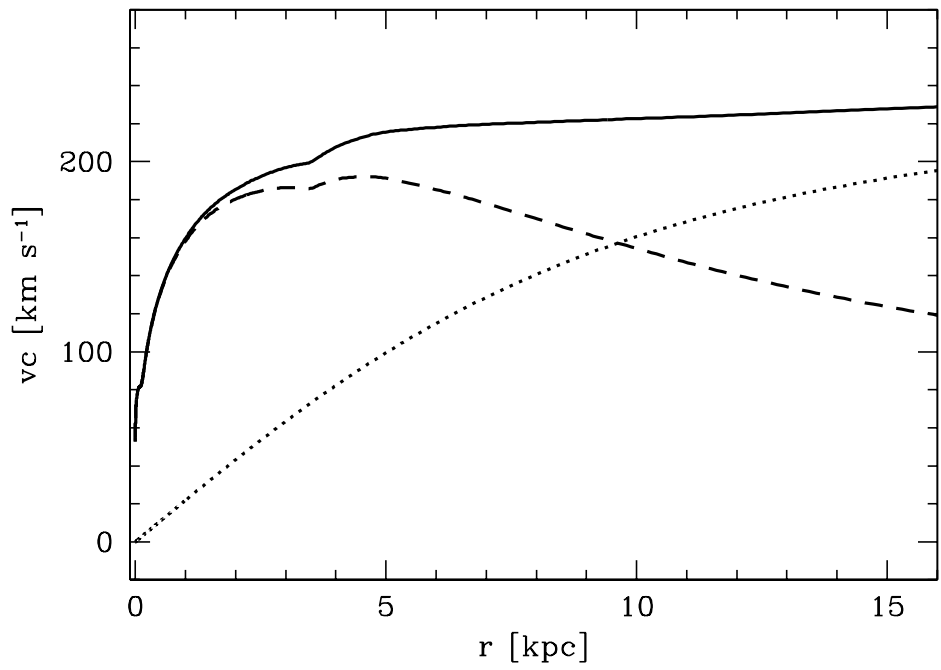}
 \caption{Rotation curve for dark matter (dotted), visible matter (dashed), and
  total (solid) for the model shown in Fig.~\ref{talk3}.
  \label{talk4}
}
\end{figure}

The combined total rotation curve can be compared to the observed
terminal velocity curve of the gas. When we assume that the luminous
matter in the Milky Way has the maximal $M/L$ allowed by the terminal
curve, and a dark matter halo scale radius of $a=10.7\,\kpc$, the best
fit is obtained with $V=235\,\kms$. The result can be seen in
Fig.~\ref{talk3}. The deviation for $r<2.5\,\kpc$ is due to the bar
which forces the gas on non-circular orbits. At larger radii, the
agreement is very good, and apart from 
two small humps at 4~kpc and 6~kpc, the model fits as well as
the scatter in the data
allows. When the orbital motion of the Sun is added to the model's terminal
velocity curve, we obtain the rotation curve in Fig.~\ref{talk4}. One sees that the
contribution of dark matter inside $R_\odot$ in the model is less than
the contribution of visible matter.

From Fig~\ref{talk3} we see that a surprisingly good fit 
to the data is obtained when
the contribution of visible matter is maximized and a constant
mass-to-light ratio is assumed. Additional evidence for this result
comes from microlensing statistics obtained for selected fields within
the bulge.  The luminous mass model is only marginally able to reach
the high microlensing optical depth measurements~\citep{bebg1997,BissantzGerhard2002}.
Increasing the dark matter component in the model would lower the
model's predictions of the microlensing optical depth, and thus increase the
discrepancies with the experimental results even more.

\begin{figure}
\includegraphics[width=0.98\textwidth]{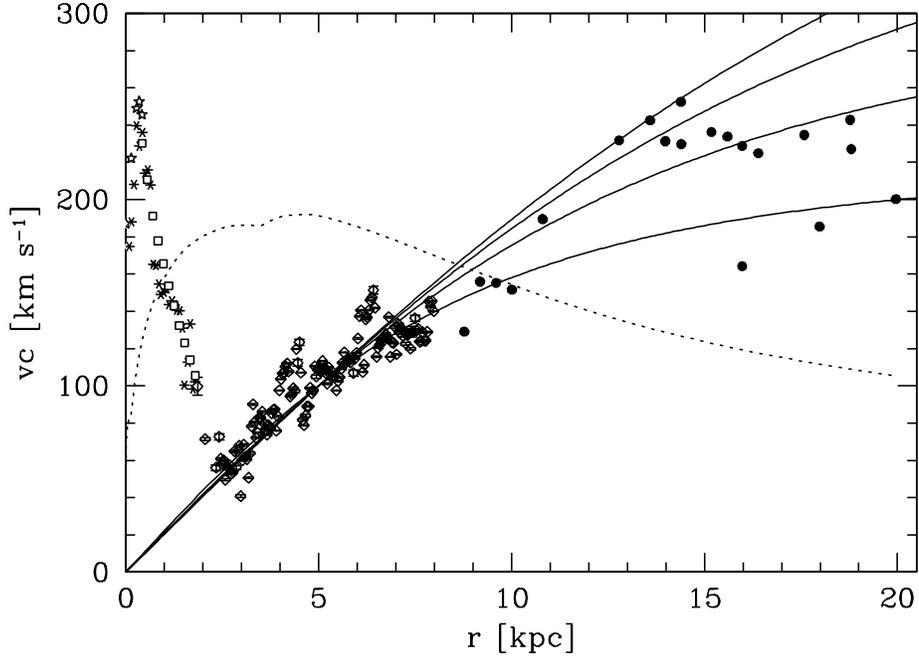}
\caption{Dark matter halo circular rotation curve deduced from
the terminal velocity curve after subtraction of the visible matter
contribution (all symbols).
The series of linearly rising solid curves represent isothermal dark
matter halo models with parameters
$(a,V)\in\{(10,223),(15,316),(20,412),(25,509)\}
(\kpc,\kms)$ and $V/a\sim21\,\kmskpc$. All these
models fit the data inside $R_\odot=8\,\kpc$ equally well. Beyond
$R_\odot$, we show data from \citet{Merrifield1992} scaled for the
values of $R_\odot$ and $V_\odot$ used here.
This figure demonstrates that data beyond $R_\odot$ is crucial for
distinguishing between different halo mass models.
The rotation curve for the visible matter is overlaid for comparison
(dotted line).
\label{talk7}
}
\end{figure}

Turning the analysis around, we can assume that we got the
contribution of the visible matter to the rotation curve right, and
then calculate the rotation curve of the dark matter halo, i.e. what
is left for the dark matter after subtraction of the known visible
matter (see Fig.~\ref{talk7}).  The data up to $R_\odot$ are taken
from Fig.~\ref{talk3}, and those beyond $R_\odot$ are taken from
\citet{Merrifield1992}\footnote{after
scaling to the values of $R_\odot$ and $V_\odot$ assumed here.}
who derived them from the HI layer thickness.
Clearly, within the solar orbit, the dark matter halo contribution to
the rotation curve rises approximately linearly with radius.  At small
radii, the data again are strongly affected by non-circular motions in
the bar potential.  The data seem to suggest that we start to see the
turnover, and that the halo scale radius is $\sim15\pm5\,\kpc$, but the
scatter in the data points beyond $R_\odot$ is quite large. Moreover,
\citet{BinneyDehnen1997} criticized that the increased scatter and
apparently linear rising rotation curve beyond the solar radius can be
explained naturally, if tracers beyond $R_\odot$ are concentrated in a
ring at $1.6 R_\odot$. If \citeauthor{BinneyDehnen1997} are correct,
this would imply that lower values for the halo scale radius $a$ are
also possible.

The analysis so far has assumed that the gas motions can be approximated by
circular orbits. However, in the bar region ($R<3\,\kpc$) this is not 
the case and at larger radii we may have significant perturbations due to
spiral arms or orbit family changes at resonances. A more detailed
analysis, however, is possible with gas dynamical models which take these
factors into account. 

\section{Orbits and resonances}

The most important orbital families in the galactic plane of a barred
galaxy are the so-called $x_1$- and $x_2$-orbits. The former exist
inside the corotation resonance (CR), where one rotation of the bar
takes as long as one circular orbit period. The $x_1$-orbits are
shaped by the periodic forcing of the bar and are in phase and
elongated with the bar.  The $x_2$-orbits are perpendicular to the bar
and typically almost round.  They exist between the inner Lindblad
resonance (ILR) and the center or between the two ILRs if
present. Lindblad resonances happen when the epicycle frequency is
twice the difference between the pattern speed and orbital angular
speed. Inside corotation we can have typically one or two ILRs;
outside corotation there is always a single outer Lindblad Resonance
(OLR). Between CR and OLR, orbits are again elongated perpendicular to
the bar but almost circular in shape.

The $x_1$-orbits are considered main building blocks of the stellar bar, 
because they support the bar potential.
They are also responsible for the sharp peak
in the Milky Way's terminal velocity curve \citep{bgsbu1991}.  

Using the atomic hydrogen and molecular gas emission all sky surveys,
which not only provide terminal velocity data for the inner galaxy,
but also information about spiral arm tangents, etc., we can use
hydrodynamical models to find further constraints for the mass model.

\section{Modeling the gas dynamics}

Given the mass model, one can find quasi steady state gas flow
solutions by assuming some equation of state for the gas and solving
the Euler equations. Here, the equation of state is assumed to be
isothermal with a sound speed of $10\,\kms$; and for the
modeling we use a 2-dimensional smooth particle hydrodynamics (SPH)
code. Artificial viscosity in the code is used to simulate shocks
properly. The background potential is split into three components:
axisymmetric, bar, and spiral arms. The bar and spiral arm components
rotate with fixed pattern speeds $\Omega_b$ and $\Omega_s$. Initially,
the gas is set up as an exponential disk on circular orbits in the
axisymmetric part of the potential. The non-axisymmetric forces due to
bar and spiral arms are introduced slowly in the model to avoid
transient features.  The resulting gas flow is stationary in the
rotating frame of the bar if bar and spiral pattern speeds are
identical, or periodic if they differ. For a given set of model
parameters, longitude-radial velocity diagrams for the gas are then
computed as would be measured by an observer rotating with the LSR,
from which we obtain the model terminal velocity curves.

\begin{figure}
\includegraphics[width=0.98\textwidth]{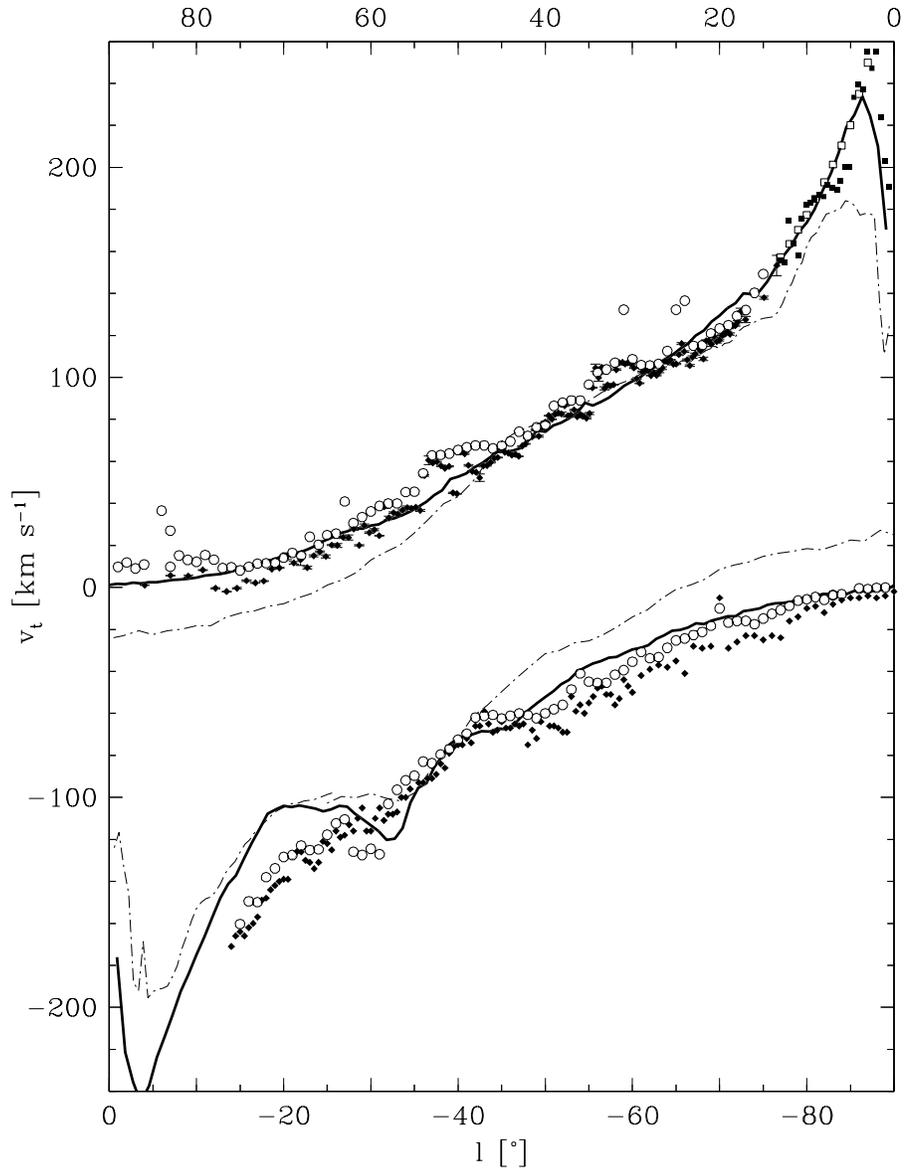}
\caption{
Gas dynamical model results for the Galactic terminal velocity curve. 
Data points are from
various observations (see \citet{EnglmaierGerhard1999} for details).
The upper (lower) part of the diagram shows data and models for
positive (negative) galactic longitudes.
The solid curve is the best model including dark
matter. The dash-dotted curve is a model without dark halo.
\label{gasterm}
}
\end{figure}

For the case of a single pattern speed, the model is evolved until the
gas flow is almost stationary\footnote{A small evolution with time is
always present because of the mass inflow towards the center.}  in the
bar frame of reference.  For the case of separate pattern speeds for
the bar and spiral patterns, the time step corresponding to the gas
flow closest to the observed gas flow is found by comparison with the
observed atomic and molecular gas kinematics.
\citet{EnglmaierGerhard1999} have shown that the terminal velocity
curve can then be well understood in the context of gas on orbits
typically found in barred galaxies. Their gas dynamical model in the
barred Milky Way model potential fits the observed terminal velocity
very well~(see Fig.~\ref{gasterm}). Note, that the high peak at
$l=2^\circ$ is explained by non-circular motions in the bar potential
and is matched better in models with higher numerical resolution.  The
whole process is iterative, because changing the position angle of the
Sun with respect to the bar major axis requires recomputation of the
mass model with the corresponding symmetry assumption, and
recomputation of the gas flow in the modified potential.

\section{The galactic bar and the spiral arms}

The best near IR model of \citet{BissantzGerhard2002} has a
3-dimensional bar with axis ratios $1:b:c$ where
$b=c=0.35\pm0.05$. The angle between the bar major axis and the line
from the Sun to the Galactic Center is $20^\circ$, with the bar's near
end at positive longitudes.

The corresponding mass distribution, however, contains no direct
information about dynamical parameters such as pattern speeds.  If we
can assume that the Milky Way's bar and spiral patterns each rotate with a
constant angular speed for at least a few rotations, and that the gas
flow reaches a quasi-steady state, then a comparison of the model with
the observed gas flow can yield constraints on the current values of
both pattern speeds in the Milky Way.

In the gas dynamical model we observe spiral arms that cause large
deviations from circular velocity within corotation and only small
deviations outside corotation of the bar. In the Milky Way, there is the
so-called '3-kpc-arm' which extends to $3\,\kpc$ in radius and has
strongly non-circular motions. On the other hand, the so-called
molecular ring material between $4\,\kpc<r<7\,\kpc$ appears to be
on nearly circular orbits. Thus we can place corotation between
those limits; in the following we set the corotation radius to $3.5\,\kpc$.

From observations and models of other barred galaxies it is known that
the corotation radius is typically $1.2\pm0.2$ times the radius of the
bar ends \citep{Athanassoula1992b}. Larger values would be possible if
the bar were able to slow down \citep{DebattistaSellwood2000}.  Our
corotation derived above is consistent with the condition $R_{\rm
CR}/R_{\rm bar}=1.2\pm0.2$, however, if the bar length turns out to be
underestimated, we would need to increase the corotation radius
accordingly.  \citet{Dehnen2000} found evidence from local stellar
streams that the OLR of the bar is located very close to the Sun,
according to \citet{MuehlbauerDehnen2003}, at $\sim0.92
R_\odot$. If so, we can calculate the corotation radius
for our model and find $R_{CR}=4.0\,\kpc$. This is in good agreement
with our previous estimate.

The gas dynamical model in the barred mass model without spiral arms
forms a 4-armed spiral pattern similar to the one deduced from
H~II-regions by \citet{Georgelin1976}; see
\citet{EnglmaierGerhard1999} for the comparison.
\citet{BissantzGerhard2002} used the more realistic spiral arm model
of \citet{OritzLepine1993} to improve the inversion of the inner
Galaxy, and \citet{beg2003} found the approximately periodic gas flow
in this modified potential, assuming separate pattern speeds, for the
bar and spiral arms.

\section{Multiple Pattern Speeds}

\begin{figure}
\includegraphics[width=\textwidth]{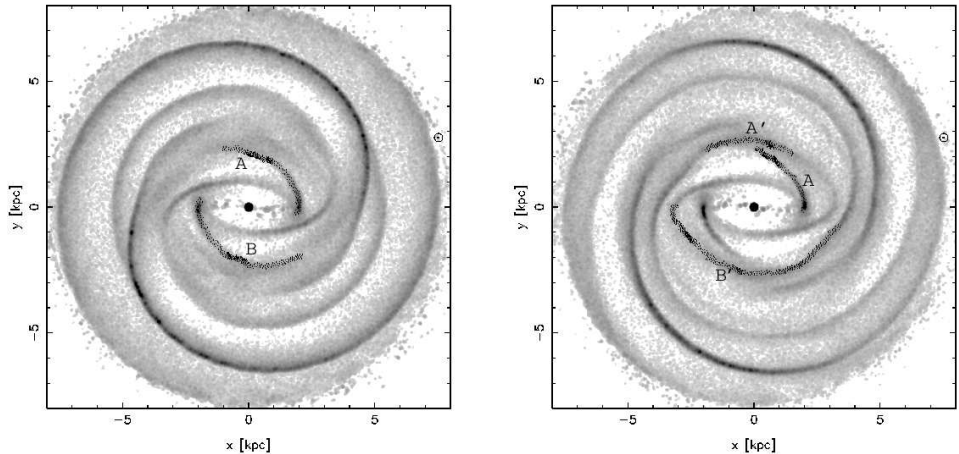}
\caption{
Gas response for models with spiral arm pattern speed set to 60 (left),
and $20\,\kmskpc$ (right).
\label{multipat1}
}
\end{figure}

\begin{figure}
\includegraphics[width=\textwidth]{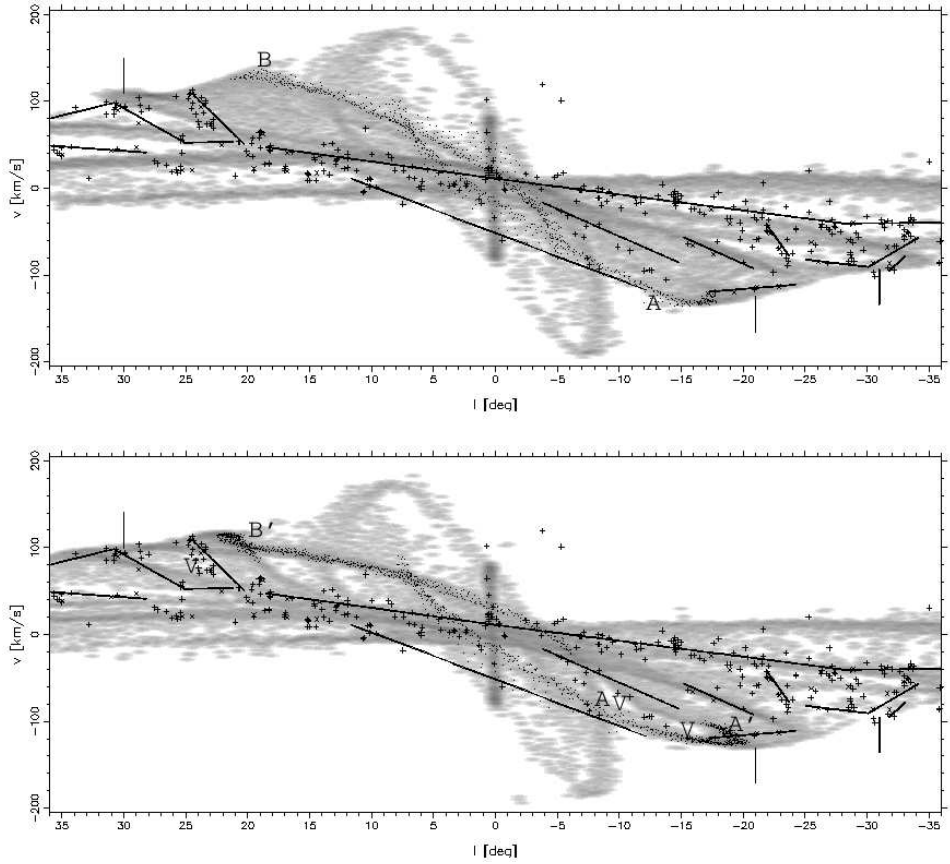}
\caption{
$(l,v)$ diagrams for models
with fast (60$\,\kmskpc$; top) and slow spiral arms ($20\,\kmskpc$;
bottom) as shown in Fig.~\ref{multipat1}. Only the slow model shows
low density voids (marked as 'V') similar to those known from observations.
\label{multipat2}
}
\end{figure}

When we allow two pattern speeds in the model, i.e. the spiral pattern
is given a different pattern speed from that of the bar, then the gas
flow is changing in a periodic fashion with the beat frequency
$\Omega_{\rm b}-\Omega_{\rm s}$.  This periodic oscillation is
strongest in the transition region between bar and spiral pattern
around the bar's corotation. Because of the lower pattern speed, the
spiral pattern extends to larger radii; its outer radius is
approximately given by its OLR.

Distant faint gaseous spiral arms have been found out to
radii\footnote{Rescaled to correct for the assumed distance of
$8\,\kpc$ to the Galactic Center.}  of 12--16~kpc \citep{Davies1972},
and even 17--22.6~kpc \citep{McClure++2004}.  However, it could well
be possible that the outermost spiral arms are transient and do not
follow the same pattern speed. Hence, it is not practical to use the
radial extent to estimate the spiral arm pattern speed.

In the model with two pattern speeds, most of the time the spiral
pattern beyond bar corotation smoothly connects to the inner spiral
pattern driven by the bar~(Fig.~\ref{multipat1}).  However, because of
the different pattern speeds, the outer spiral arms regularly separate
from the bar-driven inner arms around corotation. On the other hand,
when spiral pattern and bar rotate with the same pattern speed, the
spiral pattern weakens in the corotation region and disappears.  This
is consistent with density wave theory, which asserts that density
waves become evanescent in the region around corotation, and
decay. When two pattern speeds are used, we observe in the model that
the spiral pattern passes through the bar's corotation and through the
spiral arms' corotation radius without weakening. A detailed
comparison of the model with the observed gas flow shows that certain
gas poor regions in the observed longitude-radial velocity diagram
correspond to inter-arm regions in the gas flow near CR which only
exist in the corotation region when spiral arms and bar rotate with
different pattern speeds (Fig.~\ref{multipat2}).  The observed gas
dynamics clearly shows such gas poor regions, hence the spiral arms
must rotate more slowly than the bar \citep{beg2003}. Approximate
agreement with observations is found when a pattern speed of
$20\,\kmskpc$ is assumed for the spiral arms.

Detailed observations by \citet{NaozShaviv2005} of the spatial
separation of spiral arms and young star clusters that have moved away
from their spiral arms, also indicate not only that the spiral pattern
rotates much slower than the bar, but also that multiple pattern
speeds are present in the spiral pattern which fall into two groups
around $~18$ and $~30\,\kmskpc$. This observational result is also in
agreement with the theory of Lin \& Shu (1964) and a general result
from n-body simulations of disk galaxies, namely that spiral arms are
in fact not steady-state but recurrent and multiple modes overlap each
other (e.g. \citealt{Fux1999}).

\section{Discussion}

In this review, we gave an overview of the results obtained from
reconstruction of the present state of the Milky Way using near-IR
luminosity and gas kinematics complimentary data sets. From the
luminosity data alone, one could not infer anything about the dark
matter component. Likewise, the observed gas kinematics alone is
difficult to interpret and has lead historically to wrong conclusions,
such as steep drops in the inferred rotation curve, large dark matter
components, and explosions in the galactic center. We avoid using
parametric models for describing the important bulge region which is
best described as a barred bulge.  This bar in the galactic center
explains the observed asymmetry in the near-IR light distribution
within the bulge region as well as the high non-circular gas motion
inside 3 kpc galactocentric radius.

The inferred mass density is in good agreement with micro-lensing
optical depth observations which provide an important independent
test. The molecular ring material, is interpreted as four tightly
wound spiral arms in approximate circular rotation outside the
corotation radius of the bar. Models indicate that the spiral arm
pattern rotates much slower than the bar, allowing the arms to extend
beyond the solar orbit. Most imporantly, the model has provided
evidence for the maximum disk hypothesis and the amount of dark matter
present inside the solar radius.  Future and presently ongoing
observational surveys will provide even more detailed data which will
allow us to improve the model further. The next revision of the Milky
Way models will probably allow more detailed studies of the spiral
arms and locate more resonances.

\section*{Acknowledgments}

This work was supported by Swiss Nationalfonds grant 200020-101766.

\end{document}